\documentclass[12pt,a4paper]{article}
\renewcommand\Re{\mathop{\rm Re}\nolimits}
\renewcommand\Im{\mathop{\rm Im}\nolimits}
\begin{document}
\thispagestyle{empty}
\newcommand\order{\mathcal{O}}
\setcounter{page}{0}
\begin{flushright}
hep-ph/9702441\\
NIKHEF-97-010\\
INLO-PUB-3/97\\
February 1997
\end{flushright}

\vskip2cm

\begin{center}
{\Large\bf Light Fermion Mass Effects in $e^+e^- \to 4$~Fermions}\\[1cm]

\it Jiri Hoogland${}^\dagger{}$\\
\it NIKHEF, Postbus 41882,\\ NL-1009 DB Amsterdam,
  Netherlands\\
\tt t96@nikhef.nl\\[1cm]
\it Geert Jan van Oldenborgh${}^\dagger{}^a$\\
Instituut-Lorentz, Rijksuniversiteit Leiden,\\
  Postbus 9506, NL-2300 RA Leiden,
  Netherlands\\
\tt gj@lorentz.leidenuniv.nl\\[1cm]
\end{center}

\begin{quote}
We investigated the effect of the light fermion masses on cross
sections for $e^+e^- \to 4$ fermions in the Fermion Loop scheme
defined in Ref. \cite{BHF2}, and approximations to it.  The effects
are found to be very small, except of course in the collinear region
of single $W$ boson production where the electron mass acts as the
cut-off.
\end{quote}

\vspace{\fill}
\footnoterule
\begin{flushleft}
\footnotesize
${}^\dagger$Research supported by the Stichting FOM\\
${}^a$Current mail address {\tt oldenbor@knmi.nl}\\
\end{flushleft}
\clearpage


\section{Introduction}

Properties of the processes $e^+e^- \to 4$ fermions, which are studied
at LEP-2 and higher energy colliders \cite{LEP2report}, should be
computed using a careful treatment of the finite width of the virtual
$W$ and $Z$ bosons to obtain correct results \cite{LEP2WW}.  In Refs
\cite{stable,BHF2} a scheme was proposed that offered a solution for
the fermionic contributions to these finite width effects.  In this
`Fermion Loop scheme' all fermionic one-loop corrections to the
massive boson propagators and three-vector-boson vertices are summed
in the massless limit (except for the top quarks).  This way strict
gauge-invariance is maintained, i.e., the resulting amplitudes satisfy
all relevant Ward Identities.  As a test-case of this scheme, the
process $e^+e^- \to u \bar{d} e^- \bar{\nu}_e$ (CC20) was studied
numerically.  It was shown to behave properly in the two critical
regions: the collinear region $q_\gamma^2 = (p_e - p_e')^2 \to 0$ and
the high-energy limit $s\to\infty$.  In the first region even a small
violation of U(1) current conservation can give results which are
wrong by a factor $\mathcal{O}(m_W\Gamma_W/m_e^2) \approx 10^5$; in the
second the well-known SU(2) gauge cancellation fails when SU(2) gauge
invariance is broken.\footnote{Note that the converse is not true: the
gauge-breaking fixed-width scheme is also properly behaved for $2\to
4$ fermion processes.}

In this letter we study the effects of the inclusion of light fermion
masses in the fermion loop scheme for the CC20 process mentioned
above.  The electron mass was taken into account in the partial
calculation \cite{stable} in order to be able to take the collinear
limit, but not in the full results \cite{BHF2}.  Throughout, however,
we will neglect terms of order $m_f^2/m_W^2$ and $m_f^2/s$, as these
are much smaller than the required precision.  In principle the
inclusion of fermion masses in the Ward Identities is straightforward
but tedious.  The omission in the expressions used for numerical
implementation of some mass terms leads to small violations of these
identities; we investigate whether these have non-negligible effects.
Finally we study whether the physical effects of the light fermion
masses are observable, apart from the obvious case of the electron
mass in the forward region, where it is needed as a cut-off.

We thus follow the introduction of light fermion masses through the 
renormalization scheme, the running couplings and the cross section of the 
prototypical process $e^+e^- \to u \bar{d} e^- \bar{\nu}_e$.


\section{Renormalization}

To include the fermion loops to all orders in a compact way, it is
useful to work with the complex renormalization scheme
\cite{largewidth}, extended to the electro-weak sector of the Standard
Model.  It is straightforward to evaluate it for non-zero light
fermion masses, as all formulae in Ref.\ \cite{BHF2} are already given
for the massive case.  The results are listed in Table
\ref{tab:renormalization}; as expected the deviations are
$\mathcal{O}(m_f^2/m_W^2)$ and hence negligible.  However, at this
stage one cannot discard these yet as it might be possible that
cancellations in the amplitude will be spoiled without them.

\begin{table}
\begin{center}
$$
\setlength{\arraycolsep}{\tabcolsep}
\renewcommand\arraystretch{1.3}
\begin{array}{|l|rrr|}
\hline
\hline
m_{_W}                            &   80.10   &   80.26   &   80.42   \\
\hline
\hline
m_t                                  & 104.665   & 132.202   & 157.270   \\
\multicolumn{1}{|r|}{m_f=0} & 104.768   & 132.185   & 157.195   \\
\sqrt{\ \Re\mu_W}            & 80.074    & 80.233    & 80.393    \\
\multicolumn{1}{|r|}{m_f=0} & 80.074    & 80.234    & 80.393    \\
\Im\mu_W/\sqrt{\ \Re\mu_W}\quad & -2.0372   & -2.0503   & -2.0630   \\
\multicolumn{1}{|r|}{m_f=0} & -2.0377   & -2.0509   & -2.0636   \\
\sqrt{\ \Re\mu_Z}            & 91.1551   & 91.1549   & 91.1547   \\
\multicolumn{1}{|r|}{m_f=0} & 91.1552   & 91.1550   & 91.1548   \\
\Im\mu_Z/\sqrt{\ \Re\mu_Z}    & -2.4491   & -2.4563   & -2.4641   \\
\multicolumn{1}{|r|}{m_f=0} & -2.4538   & -2.4610   & -2.4688   \\
\hline
\Re e(m_W^2)                 &  0.311966 &  0.311978 &  0.311986 \\
\multicolumn{1}{|r|}{m_f=0} &  0.311967 &  0.311979 &  0.311986 \\
\Im e(m_W^2)                 & -0.002685 & -0.002685 & -0.002685 \\
\multicolumn{1}{|r|}{m_f=0} & -0.002685 & -0.002685 & -0.002685 \\
\Re g_w(m_W^2)               &  0.459798 &  0.460572 &  0.461396 \\
\multicolumn{1}{|r|}{m_f=0} &  0.459802 &  0.460576 &  0.461400 \\
\Im g_w(m_W^2)               & -0.006448 & -0.006482 & -0.006516 \\
\multicolumn{1}{|r|}{m_f=0} & -0.006450 & -0.006482 & -0.006516 \\
\hline
\hline
\end{array}
$$
\caption{Values of the effective top quark mass, pole positions (in GeV) and effective 
couplings at $m_W^2$ for three different $W$ masses, with massive and massless 
fermions (except top).}
\label{tab:renormalization}
\end{center}
\end{table}

\begin{figure}
\begin{center}
\setlength{\unitlength}{0.1bp}
\begin{picture}(3600,2160)(0,0)
\includegraphics{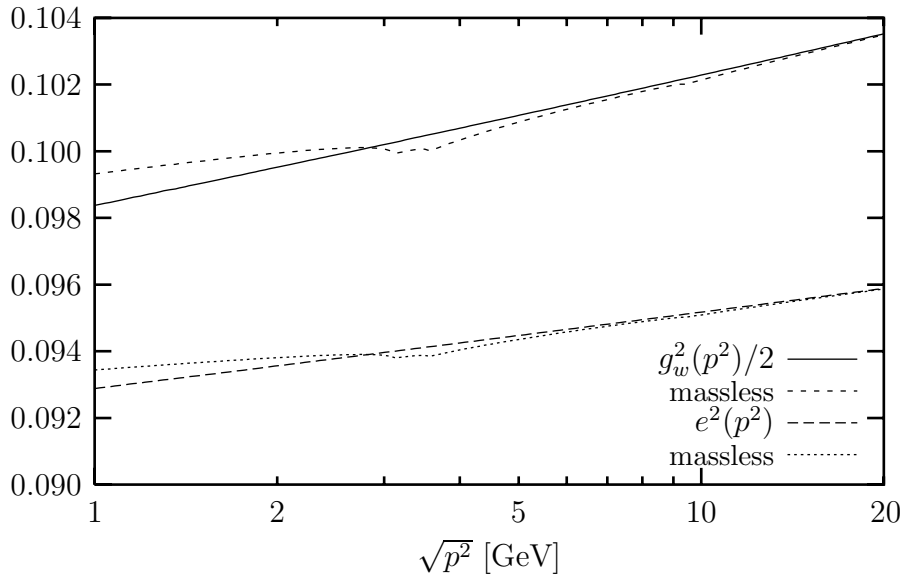}
\put(3024,413){\makebox(0,0)[r]{\small massless}}
\put(3024,533){\makebox(0,0)[r]{$e^2(p^2)$}}
\put(3024,653){\makebox(0,0)[r]{\small massless}}
\put(3024,773){\makebox(0,0)[r]{$g_w^2(p^2)/2$}}
\put(1950,30){\makebox(0,0){$\sqrt{p^2}$ [GeV]}}
\put(2749,200){\makebox(0,0){10}}
\put(3444,200){\makebox(0,0){20}}
\put(1151,200){\makebox(0,0){2}}
\put(2061,200){\makebox(0,0){5}}
\put(463,200){\makebox(0,0){1}}
\put(413,2060){\makebox(0,0)[r]{0.104}}
\put(413,1809){\makebox(0,0)[r]{0.102}}
\put(413,1557){\makebox(0,0)[r]{0.100}}
\put(413,1306){\makebox(0,0)[r]{0.098}}
\put(413,1054){\makebox(0,0)[r]{0.096}}
\put(413,803){\makebox(0,0)[r]{0.094}}
\put(413,551){\makebox(0,0)[r]{0.092}}
\put(413,300){\makebox(0,0)[r]{0.090}}
\end{picture}
\end{center}
\caption[]{The running of the couplings $e^2(p^2)$ and $g_w^2(p^2)$
at low $p^2$. 
The charm, $\tau$ and bottom quark thresholds can be seen in
the massive case.}
\label{fig:couplings}
\end{figure}

The next place where the fermion masses can play a role is in the running 
couplings defined in the fermion loop scheme.  Below the light fermion 
production threshold the slope is changed in accordance with the 
$\beta$-function, as shown in Fig.\ \ref{fig:couplings}.  One expects 
that the deviations from the massless result  (except for the top) given in 
Ref.\ \cite{BHF2} can lead to a different behaviour of the matrix element 
in the collinear (small-$q_\gamma^2$) region.

From the figure we see that even at $\sqrt{p^2}$ as low as 1 GeV the 
effects of the light fermion masses on the running couplings is less than 
1\%, and increasing only logarithmically\footnote{Note that a fit to a 
collection of straight lines would give rise to effective masses that are a 
factor 1.5 higher than the physical masses.}.  As the single-$W$ cross 
section,  which probes $e^2(p^2)$ for lower values, is suppressed relative to 
the $WW$  cross section by a factor 
$\order(\Gamma_W/m_W)\;\log(m_e^2/\mathrm{GeV}) \approx 1/10$, the effect on 
the total cross section will be small.


\section{Matrix element}

A massive tree level matrix element for the reaction $e^+e^- \to
u\bar{d}e^-\bar{\nu}_e$ was generated using MadGraph \cite{madgraph}.
This matrix element is formulated in the unitary gauge, so Higgs
ghosts do not contribute.  However, it neglects the Higgs boson
exchanges, which are suppressed by the electron mass squared.  This
matrix element originally used a gauge-invariance breaking resummation
scheme for the vector boson propagators, and hence the collinear and
high-energy behaviour were wrong.  We modified it to use the massive
version of the fermion loop scheme described below to restore the
correct collinear and high-energy behaviour.  We also added the
possibility to compare it against the ad-hoc schemes defined in Ref.\
\cite{BHF2}: naive running width, fixed width, and the approximation
of the fermion-loop scheme valid for $q_\gamma^2\to0$ (which for 
the propagators just uses the
running width).\footnote{There is one slight difference in
definitions: in the fixed-width scheme we take the pole position as
$\mu_W = m_W^2-\Gamma_W^2 - im_W \Gamma_W(1-\Gamma_W^2/m_W^2)$ to
compensate for trivial changes due to the change in definition of the
$W$ mass, which is by convention defined in a running-width scheme.}

In the latter scheme we add to the $WW\gamma$ vertex \cite{stable} an
overall factor $1 + i\Gamma_W\!/m_W\;p_+^2\!/(p_+^2 - p_-^2)$, with $p_+$ the
momentum of the $s$-channel $W^+$ and $p_-$ of the $t$-channel $W$
boson.  This minimal extension restores U(1) current conservation in
the limit $q_\gamma^2\to0$, which is sufficiently accurate for LEP2
energies.  In case of the full fermion loop scheme we implemented the
massive fermion loops as given in Appendix B of Ref.\ \cite{BHF2},
with one simplification: we assumed that the currents to which the
triangles are coupled are conserved.  In this case we get the
following expression for the tree-level and one-loop
$W(p_+^\kappa)W(p_-^\lambda)V(q^\mu)$ vertex
\begin{eqnarray}
\lefteqn{V_0^{\kappa\lambda\mu}  =  2\delta^{\mu\kappa} q^\lambda
   - 2\delta^{\mu\lambda} q^\kappa - 2\delta^{\kappa\lambda}p_-^\mu  }&&\\
\lefteqn{V_1^{\kappa\lambda\mu}  =  N_C \frac{e(q^2) g_w(p_-^2) g_w(p_+^2)}{32\pi^2} 
	\; \Biggl\{  }&&
\nonumber\\&&\mbox{}
	F_V^{(1)} \Bigl[
	p_-^\mu q^\kappa q^\lambda\; 16( 
	  - C_{22} 
	  + C_{23} 
	  + C_{33} 
	  - C_{34}
	  )
\nonumber\\&&\mbox{}
	+ \delta^{\mu\kappa} q^\lambda \; 2 (
	  - B_0(q^2) 
	  - B_0(p_+^2)
          - q^2 C_{12}
          - p_+^2 C_{12}
          + p_-^2 (C_0 + C_{12})
\nonumber\\&&\qquad\qquad\mbox{}
          - m_f^2 C_0
          - m_{f'}^2 C_0
          - 8 C_{36}
	  )
\nonumber\\&&\mbox{}
       + \delta^{\mu\lambda} q^\kappa \; 2 (
            B_0(q^2)
          + B_0(p_-^2)
          - q^2 C_0
          - q^2 C_{11}
          + p_+^2 C_{11}
          - p_-^2 (C_0 + C_{11})
\nonumber\\&&\qquad\qquad\mbox{}
          + m_f^2 C_0
          + m_{f'}^2 C_0
          - 8 C_{24}
          - 8 C_{35}
          )
\nonumber\\&&\mbox{}
       + \delta^{\kappa\lambda} p_-^\mu \; 2 (
           B_0(p_+^2)
          + B_0(p_-^2)
          + q^2 (-C_0 - C_{11} + C_{12})
\nonumber\\&&\qquad\qquad\mbox{}
          + (p_+^2 + p_-^2-2 (m_f^2+m_{f'}^2)) (C_{11} - C_{12})
          + 8 C_{35}
          - 8 C_{36}
          )
	  \Bigr] + 
\nonumber\\&&\mbox{}
	F_V^{(2)} \Bigl[
        \delta^{\mu\kappa} q^\lambda \; 2 C_{12}
      + \delta^{\mu\lambda} q^\kappa \; 2 (
            C_0
          + C_{11}
          )
       + \delta^{\kappa\lambda} p_-^\mu \; 2 (
            C_0
          + C_{11}
          - C_{12}
          )
	\Bigr] +
\nonumber\\&&\mbox{}
	F_V^{(3)} \Bigl[
       \epsilon^{\alpha\mu\kappa\lambda} {p_-}_\alpha \; 8iq^2 (
          - C_{12}
          - C_{23}
          )
       + \epsilon^{\alpha\mu\kappa\lambda} q_\alpha \; 2i (
          - 2 q^2 (C_{12} + C_{23})
          - p_+^2 C_{12}
\nonumber\\&&\qquad\qquad\mbox{}
          - (p_-^2+m_f^2-m_{f'}^2) (C_0+C_{11})
          + (m_{f'}^2-m_f^2) C_{12}
          )
	\Bigr] + 
\nonumber\\&&\mbox{}
	F_V^{(4)}  \Bigl[
	\epsilon^{\alpha\mu\kappa\lambda} {p_-}_\alpha \; 4i (
          - C_0
          - C_{11}
          + C_{12}
          )
       + \epsilon^{\alpha\mu\kappa\lambda} q_\alpha \; 4i C_{12}
	\Bigr] \Biggr\}
\end{eqnarray}
with the tensor functions $C_x(p_-^2,p_+^2,q^2)$ now in the notation of Ref.\
\cite{Passarino&Veltman}.  The coefficients are $F_V^{(i)}$ are given
by
\begin{equation}
	\left\{
	\begin{array}{rcl}
		F_\gamma^{(1)} & = & -|Q_f| \vphantom{\displaystyle |Q_f|{\strut c_w \over s_w}}\\
		F_\gamma^{(2)} & = & 0 \vphantom{\displaystyle|Q_f|{\strut c_w \over s_w}}\\
		F_\gamma^{(3)} & = & -Q_f  \vphantom{\displaystyle\frac{\strut m_f^2}{c_w s_w}}\\
		F_\gamma^{(4)} & = & 0 \vphantom{\displaystyle\frac{\strut m_f^2 I_3}{c_w s_w}}
	\end{array}
	\right.\quad
	\left\{
	\begin{array}{rcl}
		F_Z^{(1)} & = & \displaystyle |Q_f|{\strut c_w \over s_w} + \frac{1-2|Q_f|}{2c_w s_w} \\
		F_Z^{(2)} & = & \displaystyle|Q_f|{\strut c_w \over s_w} \\
		F_Z^{(3)} & = & \displaystyle\frac{\strut m_f^2}{c_w s_w} \\
		F_Z^{(4)} & = & \displaystyle\frac{\strut m_f^2 I_3}{c_w s_w}
	\end{array}
	\right.
\end{equation}
with $Q_f$ the charge of fermion $f$, $m_f$ its mass, and $c_w, s_w$ the cosine and sine of the weak mixing angle.

	Although the current attached to the photon propagator is
strictly conserved, the ones attached to the massive vector bosons are
not.  The missing terms are of order $m_f^2$, so they are negligible
if we can show that they are not scaled by a small factor (like the
naive $\order(\Gamma_W)$ gauge violation that becomes
$\Gamma_Wm_W/m_e^2$ in the collinear region).  In case of the $W$
propagators it is clear that this cannot be the case, as no
cancellation is associated with these currents, either at small or
large $p_\pm^2$.  However, there is no {\it a priori\/} reason why this
would hold for the $Z$ boson in the high-energy limit, due to the
SU(2) gauge cancellation.  Numerically, we checked {\it a
posteriori\/} that the cancellation between the three $W$-pair production
graphs in the high-energy limit is not upset by ignoring these fermion
mass terms, or by the omission of the missing Higgs-exchange terms.
All disregarded terms are thus $\order(m_f^2/m_W^2)$ and can safely be
neglected.  The cancellation is not affected by the mass terms in the 
renormalization either, which means that the renormalization (Table 
\ref{tab:renormalization}) could also be performed massless.

In order to speed up the matrix element we only call the amplitudes in
which the electron changes helicity when $q_\gamma^2 < -1000m_e^2$ and
cache the calls to the tensor functions \cite{FFguide} that will
reappear for each helicity.  With these precautions one can integrate
$10^5$ weighted points in 6 hours on a cluster of 7 Sun workstations.


\section{Cross section}

\begin{figure}[tb]
\setlength{\unitlength}{0.1bp}
\begin{picture}(3600,2160)(-200,0)
\includegraphics{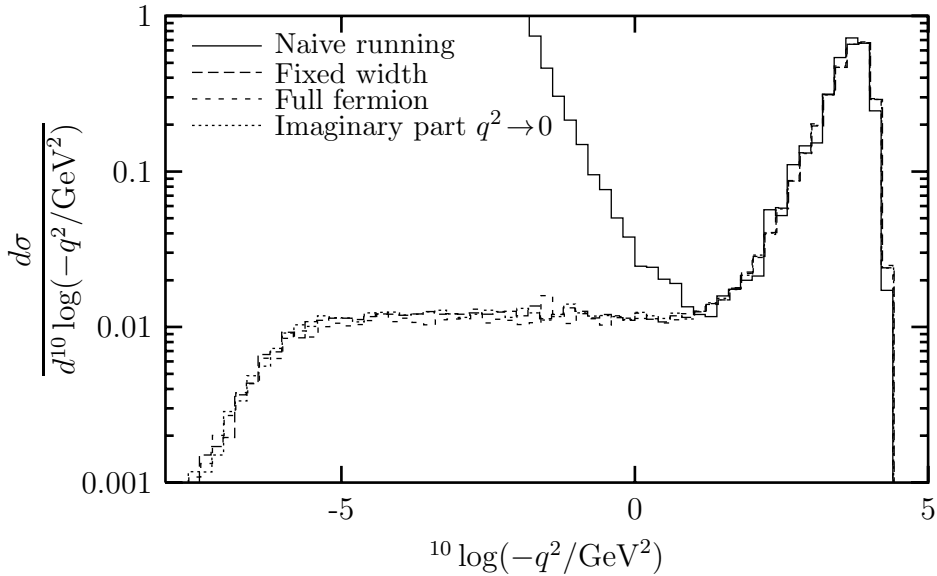}
\put(976,1647){\makebox(0,0)[l]{\small Imaginary part $q^2\!\to\!0$}}
\put(976,1747){\makebox(0,0)[l]{\small Full fermion}}
\put(976,1847){\makebox(0,0)[l]{\small Fixed width}}
\put(976,1947){\makebox(0,0)[l]{\small Naive running}}
\put(2000,20){\makebox(0,0){${}^{10}\log(-q^2/\mathrm{GeV}^2)$}}
\put(250,1180){%
\makebox(0,0)[b]{\shortstack{$\displaystyle \frac{d\sigma}{d{}^{10}\log(-q^2/\mathrm{GeV}^2)}$}}%
}
\put(3437,200){\makebox(0,0){5}}
\put(2332,200){\makebox(0,0){0}}
\put(1226,200){\makebox(0,0){-5}}
\put(513,2060){\makebox(0,0)[r]{1}}
\put(513,1473){\makebox(0,0)[r]{0.1}}
\put(513,887){\makebox(0,0)[r]{0.01}}
\put(513,300){\makebox(0,0)[r]{0.001}}
\end{picture}
\caption[]{The differential cross section for $q_\gamma^2\to0$ at $\sqrt{s}=175$ GeV of the process $e^+e^-\to u\bar{d}e^-\bar{\nu}_e$}
\label{figt}
\end{figure}

Using the matrix element described in the previous section we study
the total cross section for the reaction $e^+ e^- \to u \bar{d} e^-
\bar{\nu}_e$, at a LEP2 energy (including the forward region
$q_\gamma^2\to0$ that is sensitive to U(1) current conservation) and
at very high energies $s\to\infty$.  In order to keep the results
simple we choose not to include initial-state radiation.  First it is
shown in Fig.\ \ref{figt} that the limit $q_\gamma^2\to0$ is
well-behaved up to the kinematical limit in all schemes except the
naive running width.  In fact, the differences between the other
schemes are not significant in this plot.

The total cross section at 175 GeV, with an angle cut on the outgoing 
electron, is shown in Table \ref{tab:xs175}.  Except for the total 
cross-section without cuts, where the electron mass acts as a cut-off, there 
is no measurable difference between the massive and massless cross sections.  
Both the fixed-width scheme (with the shifted pole) and the imaginary part of 
the fermion loops in the limit $q^2\to0$ (a simple multiplicative factor) are 
seen to give results which are in good agreement.  The slightly higher values 
of the full fermion scheme result from the inclusion of part of the higher 
order corrections; the bosonic corrections would likely lower this again 
\cite{BHF2}.

\begin{table}[htb]
\begin{center}
$$
\setlength{\arraycolsep}{\tabcolsep}
\renewcommand\arraystretch{1.3}
\begin{array}{|l|r@{}lr@{}lr@{}lr@{}l|}
\hline
\theta^{\mathrm{min}}_{e^-,\mathrm{beam}}
                                    & 0^\circ&   & 0.1^\circ&  &  1^\circ&   & 10^\circ&  \\
\hline
\mbox{Running width}                & 67524&(135)\!& 1.4264&(32) &  0.6332&(7) & 0.5918&(6) \\
\mbox{Fixed width}                  & 0.6978&(10)& 0.6464&(7)  &  0.6222&(7) & 0.5924&(6) \\
\mbox{Imaginary part for $q^2\!\to\!0$} & 0.6984&(12)& 0.6456&(7)  &  0.6214&(7) & 0.5916&(6) \\
\mbox{Full fermion loops, $m_f\!=\!0$}  & \multicolumn{2}{|c}{\mbox{ill-defined}}
                                                 & 0.6527&(14) &  0.6301&(8) & 0.5988&(7) \\
\mbox{Full fermion loops, $m_f\!>\!0$}  & 0.7038&(15)& 0.6514&(9)  &  0.6298&(7) & 0.5992&(7) \\
\hline
\hline
\end{array}
$$
\caption[]{Total cross sections in pb at $\sqrt{s}=175$ GeV, without initial 
state radiation or cuts except between the outgoing electron and either beam.}
\label{tab:xs175}
\end{center}
\end{table}

\begin{figure}[htb]
\setlength{\unitlength}{0.1bp}
\begin{picture}(3600,3240)(0,0)
\includegraphics{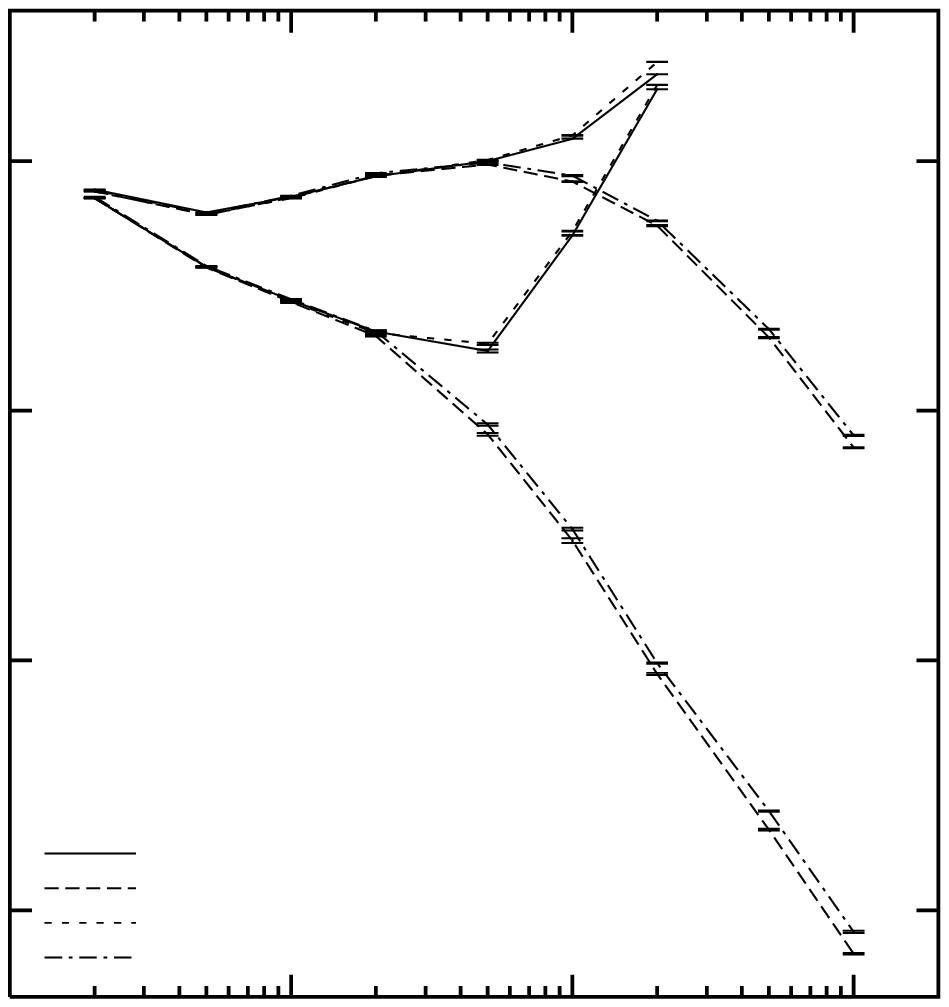}
\put(1176,413){\makebox(0,0)[l]{\small Full fermion}}
\put(1176,513){\makebox(0,0)[l]{\small Imaginary part $q^2\!\to\!0$}}
\put(1176,613){\makebox(0,0)[l]{\small Fixed width}}
\put(1176,713){\makebox(0,0)[l]{\small Naive running}}
\put(2100,0){\makebox(0,0){$\sqrt{s}$ [GeV]}}
\put(380,1720){%
\makebox(0,0)[b]{\shortstack{\large$\sigma$ [pb]}}%
}
\put(3193,200){\makebox(0,0){$10^5$}}
\put(2383,200){\makebox(0,0){$10^4$}}
\put(1573,200){\makebox(0,0){$10^3$}}
\put(763,200){\makebox(0,0){$10^2$}}
\put(713,2707){\makebox(0,0)[r]{$1$}}
\put(713,1988){\makebox(0,0)[r]{$10^{-1}$}}
\put(713,1269){\makebox(0,0)[r]{$10^{-2}$}}
\put(713,549){\makebox(0,0)[r]{$10^{-3}$}}
\end{picture}
\caption{Cross section for the process $e^+e^-\to
u\bar{d}e^-\bar{\nu}_e$ with angular cuts on the outgoing electrons
only: $1^\circ$ (upper curves) and $10^\circ$ (lower curves).}
\label{figs}
\end{figure}

Finally, we show that the inclusion of mass terms does not change the
high-energy behaviour.  As we are interested in the SU(2)
cancellations in $W$ pair production, rather than single-$W$
production, we impose a cut on the outgoing electron angle of
$1^\circ$ and $10^\circ$.  The results are shown in Fig.\ \ref{figs},
which is entirely comparable to the values in Table 5 in
Ref. \cite{BHF2}.

\section{Conclusion}

We have shown how one can incorporate the effects of fermion masses
into the fermion loop scheme that consistently resums the $W$ and
$Z$-boson propagators, up to terms of order $m_f^2/m_W^2$.  Although
we loose the rigourous proof of the Ward identities derived in Ref.\
\cite{BHF2}, we show that this does not cause problems.  Only in the
extreme forward region of phase space $q_\gamma^2 \approx -m_e^2$ does
one notice any effect of the inclusion of fermion mass terms.  The
effects of the slight shift in the renormalization scheme and the
differently running couplings are negligible.  We give the total cross
section without any cuts of the process $e^+e^-\to
u\bar{d}e^-\bar{\nu}_e$ in this scheme.

\paragraph{Acknowledgements}We would like to think the rest of the BHF
collaboration for their comments.


\end{document}